Nov 1, 2021

# A hierarchical levitating cluster containing transforming small aggregates of water droplets

Alexander A. Fedorets [a], Leonid A. Dombrovsky [a,b,c], Edward Bormashenko [c], and Michael Nosonovsky [a,d, 1]

[a] X-BIO Institute, University of Tyumen, 6 Volodarskogo St, Tyumen, 625003, Russia.
Tel. +7-345-259-7425, fedorets_alex@mail.ru

[b] Joint Institute for High Temperatures, 17A Krasnokazarmennaya St, Moscow, 111116, Russia.
Tel. +7-910-408-0186, ldombr@yandex.ru

[c] Department of Chemical Engineering, Biotechnology and Materials, Engineering Science Faculty, Ariel University, Ariel, 40700, Israel. Tel.: +972-074-7296863, edward@ariel.ac.il

[d] Department of Mechanical Engineering, University of Wisconsin–Milwaukee, 3200 North Cramer St, Milwaukee, WI 53211, USA, Tel. +1-414-229-2816, nosonovs@uwm.edu

**Abstract**

A new type of a levitating droplet clusters composed of often transforming small aggregates of water droplets is described for the first time. Unlike earlier observed droplet clusters controlled by aerodynamic forces, which formed either an ordered hexagonal structure or a chain structure, the cluster under consideration has a hierarchical organization. Small groups of closely spaced or packed droplets with interactions controlled by the electrostatic force are combined into larger structures controlled by aerodynamic forces. Since charged droplets in the nucleus of the cluster do not have dynamically stable configurations, droplet aggregates keep continuously restructuring. However, droplets with lower charge in external layers of the cluster form a stable hexagonal structure.

**Keywords**:  droplet cluster, aggregates of droplets, levitation

---

[1] Corresponding author. Tel. +1-414-229-2816, nosonovs@uwm.edu

**Introduction**

Close-packed colloidal crystals made of rigid microparticles are a natural phenomenon which has been studied for decades. Cooperative motion, self-assembly of ordered structures and structural phase transitions in colloidal crystals are similar to effects observed in condensed matter [1]. Recently, small colloidal clusters have attracted attention of researchers. Such clusters of several particles to several dozens of particles can form close-packed structures while levitating due to acoustic waves or due to another mechanism. Small clusters often possess properties absent from large crystals.

Unlike close-packed structures with neighboring particles touching each other, levitating particles can form configurations and possess symmetries not typical of large crystals. Perry et al. [2] investigated levitating 2D spherical particles of solid sulfate polystyrene which formed clusters demonstrating structural rearrangements. In certain configurations, individual particles in these clusters may have bonds between them due to aerodynamic forces. For a system of six particles, Perry et al. [2] identified seven-bond and eight-bond configurations. Lim et al. [3] reported transitions between sticky and ergodic configurations in similar six-particle and seven-particle systems which could form various arrangements, with different characteristic probabilities likely associated with the Zipf probability distribution [4].

Self-assembled clusters of condensed microdroplets levitating over a locally heated water surface have been first reported by Fedorets in 2004 [5]. When a thin (less than 1 mm thick) layer of water is heated locally to the temperatures at which water evaporates actively, small water droplets (usually from 5 μm to 50 μm in radius) are condensed in the ascending flow of mixed air and water vapor. Droplets tend to form monolayer clusters levitating at height comparable with their radii. Such clusters are typically arranged into an ordered hexagonal structure due to an interplay of aerodynamic forces which drag the droplets towards the center of the heated flow facilitating closed packing and aerodynamic repulsion forces between droplets facilitating a distance between them. Besides large (dozens to hundreds of droplets) hexagonally ordered droplets, small clusters (from one to dozens of droplets) have been discovered and a methodology to synthesize them has been developed. Small clusters may possess symmetries absent from the large clusters including the 4-fold, 5-fold, and 7-fold symmetries [6]. It has been hypothesized that the unusual symmetries can be classified with the simply laced Dynkin diagrams (the ADE-classification) [7]. Small clusters demonstrate collective behavior including simultaneous horizontal and vertical oscillations [8-9].

Besides the clusters with hexagonal arrangement of droplets, chain-like arrangement was found under certain conditions. The transition between the hexagonal and chain arrangements is reversible and it possesses characteristics of a phase transition [10]. When the thermocapillary flow is not suppressed by surfactants, a ring cluster surrounding the convective vortex flow can also form [11].

Here we report a new type of cluster arrangement, hierarchical clusters (H-clusters). These are relatively large clusters consisting of small groups or aggregates of droplets with bonds between individual droplets in each group. Unlike in previously studied clusters, droplets in the hierarchical cluster may carry electrical charge and electrostatic forces play a role in droplet group arrangement, while aerodynamic forces control interactions between the droplets as usual [12].

**Experimental results**

The droplet clusters were formed from microdroplet clouds created by an ultrasound nebulizer, which generated water microdroplets by ultrasonic acoustic irradiation. Ultrasonic nebulization of water and other polar liquids results in electrically charged microdroplets (the Lenard or ballo-electric effect). In a dry environment, the droplets keep their charge for a significant time. In particular [13], both positively and negatively charged water droplets (the latter constitute about 10% of all droplets) are formed by ultrasonic nebulization. The typical charge is on the order of hundreds or thousands of elementary charges with the probability density function decreasing quickly with increasing charge. Droplets with the charge of dozens of thousands of the elementary charge are also exist but they are rare [13].

A typical lifecycle of the H-cluster is presented in **Fig. 1**. Similarly to earlier reported chain clusters [10], the H-clusters form when droplets become large enough, so that vortex flows in between the droplets facilitate the structure formation. Such vortices are observed in some of the videos showing small particles on water surface. While weakly charged droplets forming chains are condensed in humid air, those formed in dry air result in H-clusters, which are the focus of the present study. For the H-clusters, complex electrostatic interactions between droplets are imposed over complex aerodynamic forces generated by the gas flow. Consequently, clusters are built not just of single droplets, but of small groups of droplets (typically two to four), which we call L-groups.

At first, many charged droplets penetrate into the cluster (Fig. 1a) causing rapid merging of droplets and coalescence of individual droplets with the water layer (which does not cause the collapse of the entire cluster [14]). These processes led to the decrease in the total number of droplets in the cluster, *N*. Thus, in forty seconds separating the snapshots in Fig. 1(a) and Fig. 1(b), the number of droplets decreased by approximately one third of the initial number of droplets. As the number of charged droplets decreases, the intensity of droplet merging and coalescence with the substrate decreases as well, although this process go on during the entire life cycle of the H-cluster.

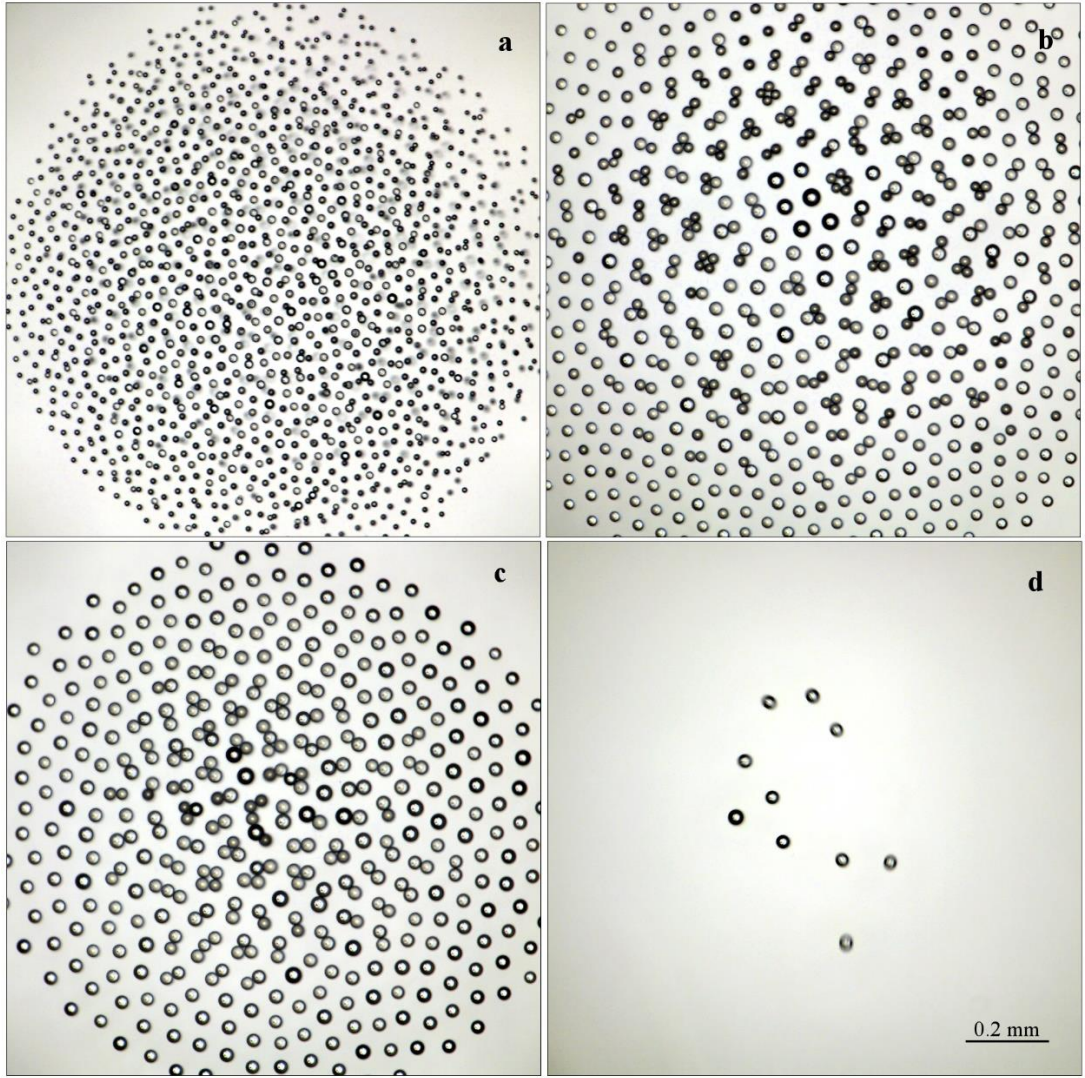

Figure 1. Lifecycle stages of the H-cluster (recorded at 5 frames per second):
a – $t = t_0$, $N > 1200$; b – $t = t_0+40$ s, $N > 800$; c – $t = t_0+63$ s, $N = 437$; d – $t = t_0+63.2$ s, $N = 10$.

High frequency video recordings show some details of the merging and coalescence (**Fig. 2**), which occurs at the time on the order of several microseconds. At frequencies of 50 frames per second the coalescence of two droplets looks like disappearance of the two droplets and emergence of a new large droplet (**Fig. 2c,d**), while the coalescence with the layer looks like disappearance of the droplet (**Fig. 2a,b**). In some cases, the merging of two or even three droplets cause also the coalescence of the larger droplet with the layer. Thus, **Fig. 2c,d** shows coalescence of a large droplet 0.12 s after it was formed by merging of three small droplets. However, when small droplets merge at the earlier stage of cluster's evolution, the larger droplets remain stable for dozens of seconds. These larger droplets are seen in **Fig. 1b**. In some cases, almost simultaneous merging of more than two droplets occurs.

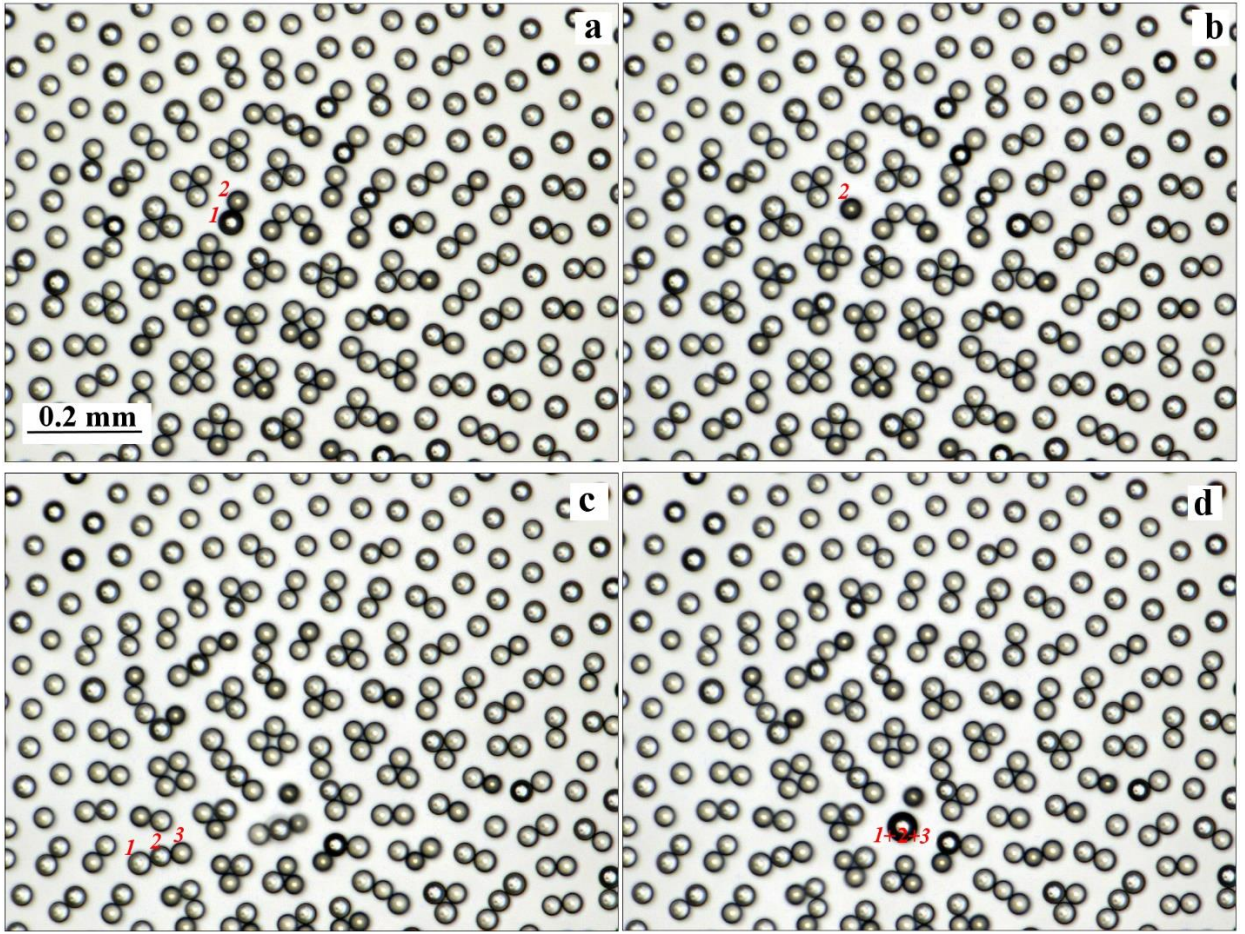

Figure 2. From (a) to (b) – coalescence of a droplet with the water layer in the time interval between $t = t_1$ and $t = t_1+0.02$ s; from (c) to (d) – merger of three droplets into one between $t = t_2$ and $t = t_2+0.02$ s (recorded at 50 frames per second).

The analysis of a series of frames shows that at the initial stage of H-cluster formation, the number of single droplets decreases sharply (Table 1). Most common groups are of four or less droplets, with a statistical distribution similar to the Zipf law.

Table 1. Frequency distribution of groups of various size at different times (s).

| | Type | $t = 0$ | $t = 1$ | $t = 2$ | $t = 3$ | $t = 4$ | $t = 5$ |
|---|---|---|---|---|---|---|---|
| A | ○ | 32 | 24 | 19 | 9 | 13 | 10 |
| B | ∞ | 36 | 35 | 37 | 41 | 36 | 34 |

| C | ⚪⚪⚪ | 16 | 11 | 9 | 10 | 13 | 13 |
|---|---|---|---|---|---|---|---|
| D | ⚪⚪⚪⚪ | 2 | 5 | 7 | 17 | 7 | 8 |
| E | ⚪⚪⚪⚪ | 2 | 5 | 3 | 3 | 5 | 3 |
| F | ⚪⚪⚪⚪ | 2 | 0 | 2 | 2 | 1 | 1 |
| G | ⚪⚪⚪⚪ | 2 | 2 | 0 | 2 | 0 | 0 |
| H | ⚪⚪⚪⚪ | 0 | 1 | 0 | 0 | 0 | 0 |
| I | ⚪⚪⚪⚪ | 0 | 0 | 0 | 1 | 0 | 0 |
| J | ⚪⚪⚪⚪⚪ | 0 | 1 | 0 | 0 | 0 | 1 |
| K | ⚪⚪⚪⚪⚪ | 0 | 0 | 0 | 0 | 0 | 1 |
| L | ⚪⚪⚪⚪⚪ | 0 | 0 | 0 | 0 | 1 | 0 |
| Total groups | | 92 | 84 | 77 | 85 | 76 | 71 |
| Total droplets (in the nucleus) | | 182 | 181 | 161 | 205 | 177 | 170 |

Note that different droplets may look differently at the same photographs with some of them having a light "nucleus" (**Fig. 3**). This is caused by different levitation height of the droplets, thus droplet 1 is above droplet 2, and their intersecting is observed, constituting at its maximum 10% of the diameters (**Fig. 3**). One can conclude that the difference in their levitation heights constitute about one radius. Similarly, heavy droplets close to the layer surface due to their increased mass have a light nucleus (droplets 3, 4, 5 in **Fig. 3**). We will denote lower levitating droplets as type "n" and higher levitating droplets as type "p". Thus in the same cluster different droplets can levitate at significantly different heights. This is another indication of the presence of differently charged (by sign) droplets in the H-cluster.

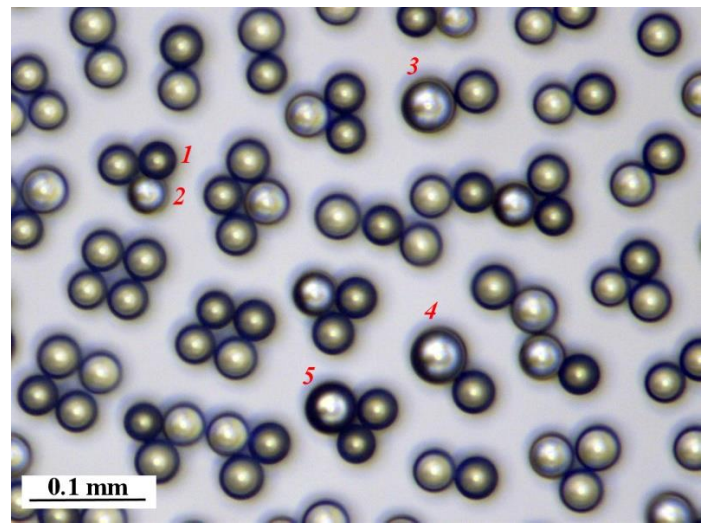

Figure 3. Classification of droplets based on the levitation height.

It is known that droplets tend to attain a small positive charge during their condensational growth [12]. The surface of the evaporating water layer can charge positively. Consequently, an electrostatic interaction force between the droplet and the layer is added to the aerodynamic drag force of the ascending gas flow. The equilibrium levitation height of positively charged droplets grows, while for negatively charged droplets the levitation height decreases down to the coalescence with the water layer.

The observed property of the L-groups is that droplets are continuously moving the groups are restructuring exchanging droplets and fragments of several droplets. The type of the droplet can change as well, as evidence by rapid video-recording. Thus, it is observed that a group of five droplets is divided into groups of two and three droplets within dozens of milliseconds. The droplet 2 changes its n-type into the p-type, while droplet 3 changes p-type for n-type (**Fig. 4**). This indicates that the type of the droplet at any instance of time depends on its neighbors.

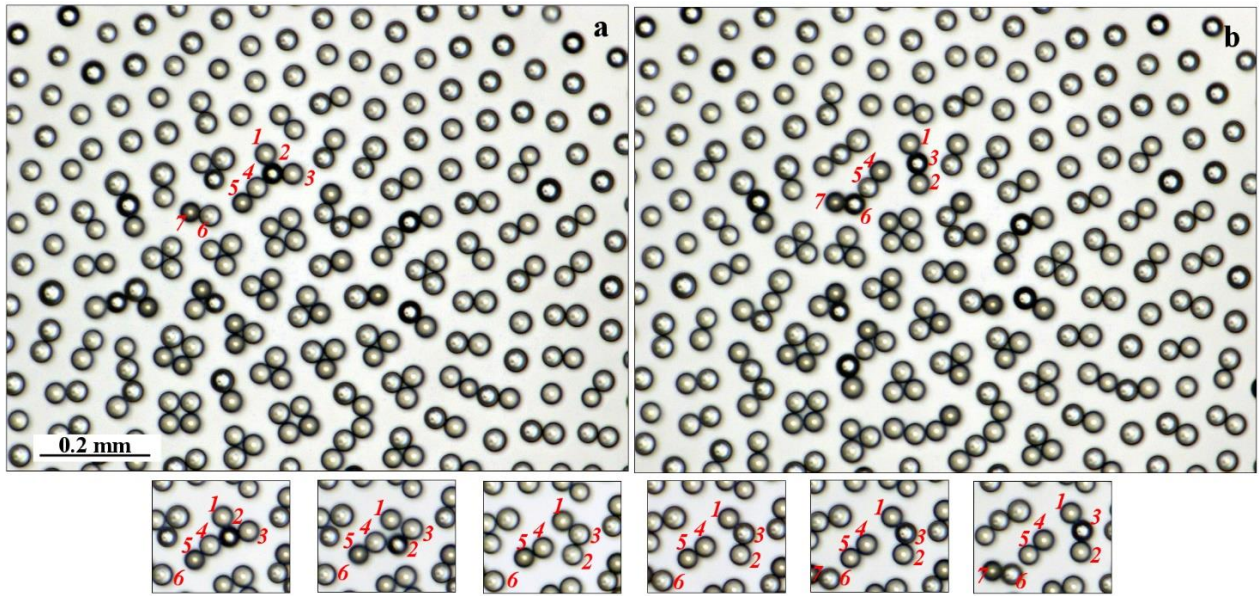

Figure 4. Rearrangement of the L-groups of water droplets in the time interval from $t = t_0$ (a) to $t = t_0+0.07$ s (b). The intermediate time moments $t = t_0+0.01j$ ($j = 1,…,6$) are presented in the lower small images.

**Discussion**

The H-clusters are formed with the use of the nebulizer and they emerge only in dry air. The relative humidity of the ambient air in the lab was at the levels not exceeding 8–12%. At higher levels of the relative humidity, the charge is lost quickly and a usual hexagonal cluster is formed. The level of humidity determines whether the electric charge relaxes from the droplets.

Note that droplets cannot be treated as point charges, they clearly have a dipole moment. Due to the large fraction of initially weakly charged droplets in the aerosol and the loss of charge, only some droplets in the cluster participate in the formation of L-groups. Thus, at the final stage (**Fig. 1c**), only 177 droplets out of the total $N=437$ (i.e., about 40 %) participate in L-groups.

The structure and dynamic behavior of the central part of the H-cluster can be explained assuming that the droplets carry an electrical charge, either positive or negative. The charge results into the electrostatic forces comparable with the aerodynamic forces acting upon a regular droplet cluster in the absence of the electric charge. However, the aerodynamic forces still dominate. Unlike in a regular cluster of uncharged droplets, which has a relatively stable hexagonal equilibrium configuration (as observed at the periphery of the H-cluster), the central part of the cluster does not possess the equilibrium configuration. On the contrary, one can see continuous rearrangement of the droplets and their groups.

In order to attain the qualitative understanding of the effect lets us separate the characteristic scales, inherent for the hierarchical droplet cluster. Let us start from the "small scale", which is close to the radius of the droplet, on which non-coalescence of the droplets, separated by the distance which is much smaller than the radii of the droplets, is observed. What is the reason for this non-coalescence? Obviously, the coalescence diminishes the interfacial energy of droplets, resulting in the effective attraction force, acting between the droplets. Let us estimate the characteristic scale $R^*$ at which electrostatic effects become comparable to the interfacial ones. Compare the energy electrostatic interaction between two charged droplets possessing electrical charge $Q$ and radii $R$ separated by the negligible distance to the interfacial energy of the droplets:

$$k\frac{Q^2}{2R} \cong 8\pi\gamma R^2$$

The length scale at which these energies become comparable are given by

$$R^* \cong \sqrt[3]{\frac{kQ^2}{16\pi\gamma}}$$

The realistic estimation is $Q \cong 1000e = 1.6 \times 10^{-16}$ C, substitution this value and assuming $\gamma = 70 \times 10^{-3} \frac{J}{m^2}$ yields $R^* \cong 1.0 \times 10^{-7}$ m $= 100$ nm. For the droplets larger than $R^*$ (and this is the case in our experiments) interfacial effects will dominate on the electrostatic ones, and interfacial attraction will prevail upon the electrostatic repulsion. Thus, electrostatic repulsion is definitely not responsible for the non-coalescence. It seems plausible to relate the effect of non-coalescence to the rotation of droplets or convective motion of water neat the surface of non-isothermal droplets [14] drawing thin air layer adjacent to the surface of droplets, as shown in [15]. It was demonstrated that the presence of explicitly moving interfaces may provide a lubrication pressure which can resist coalescence for macroscopically long times. Thus, from the observation of a cluster with fluorescent particles embedded in its droplets [16], it is found that a typical revolution period of a droplet is on the order of 1 s.

Both the droplet rotation and water flow along the surface inside the droplet are accompanied by formation of a thin layer of a moving gas in the gap between the neighboring droplets of L-groups. In this gap, the gas velocity varies linearly between the local velocities of water at the surfaces of two water droplets. The gas layer works like a kind of lubricant that promotes the continuation of the almost independent rotation of the droplet (or the water flow under their surface). The small value of the dynamic viscosity of a gas ensures the high quality of such "gas lubrication". Judging by the rare coalescence of the droplets in L-groups, complete overlap of the gas gap in a group of closely spaced droplets rarely occurs. This means that maintaining a thin gas gap between the droplets is a typical situation for the L-groups, explaining the observed frequent rearrangement of droplets in L-groups. It is also instructive to estimate the length scale at which the aerodynamic interaction between droplets become comparable to the interfacial effects. The aerodynamic attraction between the droplets may be roughly described with the potential [17]:

$$U_{aer}(r) \sim \xi \frac{4\pi^2}{3\sqrt{3}} \rho u_0^2 \frac{R^4}{r} \quad ,$$

where $R$ is the radius of the droplet, $u_0$ is the characteristic velocity of the steam, $\xi < 1$ is the dimensionless coefficient and $r$ is the separation between the droplets. The effects due to aerodynamic attraction become comparable to interfacial ones when the following equation takes place:

$$\xi \frac{4\pi^2}{3\sqrt{3}} \rho u_0^2 \frac{R^4}{r} \cong 8\pi \gamma R^2$$

The length scale at which aerodynamic and interfacial effects become comparable is given by:

$$r^* = \frac{\xi \pi}{6\sqrt{3}} \frac{\rho u_0^2 R^2}{\gamma}$$

Assuming $\xi \cong 1, \rho = 0.65 \frac{kg}{m^3}, u_0 = 0.08 \frac{m}{s}, R = 20\ \mu m, \gamma \cong 70 \times 10^{-3} \frac{J}{m^2}$ yields the unreasonable estimation $r^* \sim 10^{-11}$m. This means that the coalescence of droplet is driven by interfacial and not be aerodynamic effects, and consequently the aforementioned air lubrication effect have to withstand the interfacially driven attraction of droplets.

Now consider the large scale (~10mm) effects. We already demonstrated that on the large-scale aerodynamic effects, contrastingly, prevail on electrostatic ones [17].

To estimate the ratio of the electrostatic and aerodynamic forces and their characteristic length scales, one can consider the electrostatic force between the droplets. Note, however, that droplets cannot be viewed as point charges but rather than as dipoles. Consequently, the characteristic size of the electrostatic attraction force is small in comparison with the droplet size, so it is significant only within a group. The aerodynamic force has much larger characteristic distance? so it applies to the entire cluster rather than a small group.

The electric charge is created due to the Lenard (or ballo-electric) effect during the dispersion of liquid. Keeping of the droplets electrical charge is very sensitive to the ambient humidity, since the charge remains only in dry air. In humid air, the electric charges are lost, in particular those of the positive sign, as condensational growth of the droplets often creates negative charges.

Unlike the H-cluster, the chain cluster is controlled by pure aerodynamic origin, and it is observed when droplets are electrically neutral. The interplay of aerodynamic and electrostatic forces creates any effects in the H-cluster which are of potential interest for the 2D aerosol technology.

The H-cluster can be characterized as chaotic or unstable, because it keeps rearranging. At the same time, it demonstrates certain orderliness, although it is less ordered than the hexagonally ordered droplet cluster and the chain droplet cluster. The H-cluster can therefore be viewed as a new type of colloidal crystal, which is chaotic but still ordered.

**Materials and methods**

*The experimental setup* is presented in **Fig. 5**. The cluster (1) was formed over a locally heated spot of a horizontal layer of distilled water (2) containing surfactant microadditives (sodium dodecyl sulfate) suppressing thermocapillary flows. A sitall substrate (400 μm thick) was attached to the bottom of the cuvette (4). The body of the cuvette contained channels connected to the cryothermostat CC805 (Huber, Germany) allowing temperature stabilization in the water layer. Local heating of the water layer can be performed with a laser beam (MRL-III-660D-1W, CNI, China) (5) directed towards the lower side of the substrate. (6) is the camera lens.

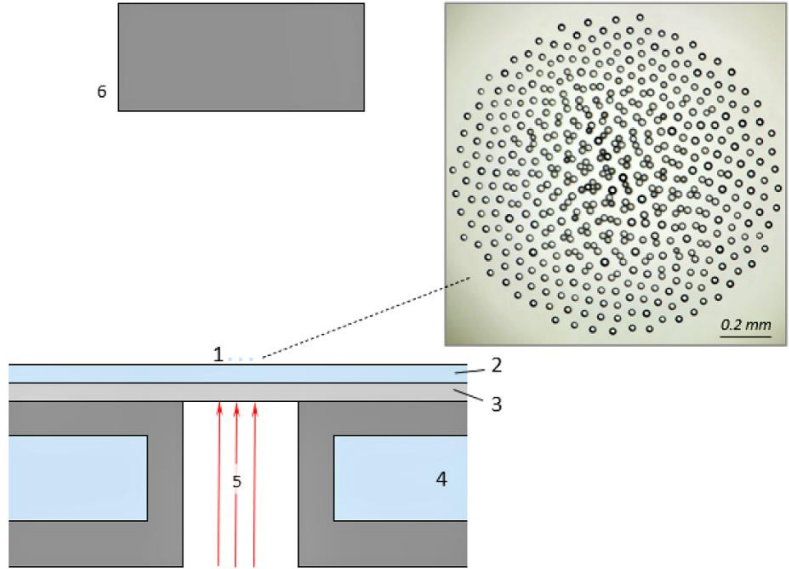

Figure 5. Schematics of the experimental setup (side view) and a typical H-cluster (top view).

Temperature field on the water layer surface was detected by an IR camera A655sc (FLIR, USA). Visual recording of the cluster was conducted with a stereomicroscope AXIO Zoom.V16, (Zeiss, Germany), with the attached high-speed camera PCO.EDGE 5.5C (PCO, Germany). In all experiments the thickness of the water layer was maintained at $400 \pm 2$ μm, and thermostatic temperature $T_0 = 15 \pm 1$ °C. The cluster was generated in an open cuvette, with the ambient air temperature in the lab $26 \pm 1$ °C.

*Synthesis of H-clusters.* There are two requirements for the H-cluster synthesis. *First*, the H-cluster is formed from a cloud of microdroplets generated by an ultrasonic nebulizer. Second, the ambient air in the laboratory must be very dry: the relative humidity is about ten percent or less.

The microdroplet cloud was created by an ultrasound nebulizer (Omron, Japan), which generated aerosol microdroplets (diameter about 5 μm) by ultrasonic acoustic irradiation. The aerosol was injected when the heating power level was decreased (the local temperature of the water layer at the center of the heating spot was $T_{max} = 60 \pm 2$ °C), which facilitated the penetration of small droplets to the cluster area. It was also important to keep the density of droplets at the heating spot is high. After that, the heating power was abruptly increased with temperature stabilized at 75 ± 2 °C, and an active condensational growth of the droplets as well as their coalescence occurred. As a consequence of the growth of the droplet sizes, at a certain point, various local multi-droplet structures started to form. With further increasing size of the droplets, the cluster collapses due to coalescence of the droplets.

**Acknowledgements.** The authors gratefully acknowledge the Russian Science Foundation (project 19-19-00076) for the financial support of this work.